\documentclass[a4paper,10pt]{article}

\usepackage{amsthm}
\usepackage{amsmath}
\usepackage{amssymb}

\usepackage{tikz}
\usepackage{enumerate}
\usepackage{setspace}

\usepackage{caption}
\usepackage{chngcntr}
\counterwithout{figure}{section}

\usepackage{a4wide}
\usepackage{hyperref}
\newcommand{\Up}{U_+}

\newcommand{\Con}{\ensuremath{\mathcal{C}}}

\newcommand{\Cinf}{\ensuremath{\mathcal{C}^\infty}}

\newcommand{\loc}{\ensuremath{\text{loc}}}

\newcommand{\mb}[1]{\ensuremath{\mathbb{#1}}}

\newcommand{\R}{\mb{R}}

\newfont{\bl}{msbm10 scaled \magstep2}
\newcommand{\beq}{\begin{equation}}
\newcommand{\eeq}{\end{equation}}

\renewcommand{\d}{\ensuremath{\partial}}

\newcommand{\pder}[2]{\frac{\d#1}{\d#2}}

\newenvironment{pr}{\begin{proof}[\textbf{Proof:}] \ }{\end{proof}}
\newtheorem{thm}{Theorem}[section]

\newtheorem{ex}[thm]{Example}
\newtheorem{cor}[thm]{Corollary}
\newtheorem{rem}[thm]{Remark}
\newtheorem{defi}[thm]{Definition}

\newcommand{\ep}{\epsilon}

\numberwithin{equation}{section}

\newcommand{\dd}{{\mathrm{d}}}
\def \d {{\mathrm{d}}}
\newcommand{\bW}{{\bf W}}
\newcommand{\bF}{{\bf F}}
\newcommand{\bE}{{\bf E}}
\newcommand{\bP}{{\bf P}}
\newcommand{\bQ}{{\bf Q}}
\newcommand{\bT}{{\bf T}}

\def \H {{\mathcal H}}

\def \U {{\cal U}}
\def \V {{\cal V}}

\title{The global existence, uniqueness and $\mathcal{C}^1$-regularity of geodesics in nonexpanding impulsive gravitational waves}
\author{
J.~Podolsk\'y$^1$\thanks{{\tt podolsky@mbox.troja.mff.cuni.cz}},
C.~S\"amann$^2$\thanks{{\tt clemens.saemann@univie.ac.at}},
R.~Steinbauer$^2$\thanks{{\tt roland.steinbauer@univie.ac.at}}
and R.~\v{S}varc$^1$\thanks{{\tt robert.svarc@mff.cuni.cz}} \\ \\
$^1$ Institute of Theoretical Physics,\\
Charles University in Prague, Faculty of Mathematics and Physics, \\
V Hole\v{s}ovi\v{c}k\'ach 2, 18000 Prague 8, Czech Republic.\\ \\
$^2$ Faculty of Mathematics, University of Vienna, \\
Oskar-Morgenstern-Platz 1, 1090 Vienna, Austria. \\ \\
}
\date{October 31, 2014}

\begin{document}

\maketitle

\begin{abstract}
We study geodesics in the complete family of nonexpanding impulsive
gravitational waves propagating in spaces of constant curvature, that is Minkowski, de~Sitter and anti-de~Sitter universes.
Employing the continuous form of the metric we prove existence and uniqueness of
continuously differentiable geodesics (in the sense of Filippov) and use a
$\mathcal{C}^1$-matching procedure to explicitly derive their form.
\end{abstract}

\section{Introduction}
Impulsive gravitational waves describe short but intense bursts of gravitational
radiation and have become physically interesting models of exact radiative
spacetimes in Einstein's theory. At the same time they are spacetimes of low
regularity: While one prominent form
of the metric is only continuous (actually locally Lipschitz continuous --- a
fact that will turn out to be essential in our
analysis) another one is even distributional. Hence these geometries are also
mathematically interesting in the context of non-smooth Lorentzian geometry,
where they can
serve as relevant test models.

In this work we treat the \emph{entire class of nonexpanding impulsive waves}
propagating on \emph{spaces of constant curvature} --- Minkowski space, de Sitter
and anti-de Sitter universes (with vanishing, positive and negative cosmological constant $\Lambda$, respectively). 
We focus on particle motion using the
continuous form of the metric. In particular, we prove that the geodesics are
continuously differentiable curves and then we apply a $\mathcal{C}^1$-matching
procedure to compute them explicitly.

Specifically, we begin in section \ref{sec:igw} with a review of the class
of nonexpanding impulsive gravitational waves in spaces of constant curvature, including
various methods of their construction. In section~\ref{sec:geo} we
focus on the main topic of this work: the geodesic equation in the
continuous form of the metric. After reviewing in sections \ref{sec:3.1} and \ref{sec:3.2} what has 
been done so far we derive in section \ref{sec:3.3} 
our key mathematical result, a general existence and uniqueness theorem
for the geodesic equation in a class of locally Lipschitz continuous spacetimes. For this purpose we
use \emph{Filippov's solution concept }for differential equations with
discontinuous
right hand side whose basics are collected in appendix~\hyperref[app:a]{A}.
After explicitly deriving the geodesic equations in section~\ref{sec:3.4}, 
in sections \ref{mainlz} and \ref{mainlnz} we apply these mathematical findings to establish our
main result (for ${\Lambda=0}$ and ${\Lambda\not=0}$, respectively), which says that the entire class 
of nonexpanding impulsive waves has unique, continuously differentiable global geodesics. In section
\ref{sec:matching} we explicitly derive the geodesics for this class of
spacetimes using a $\mathcal{C}^1$-matching procedure.

\section{Impulsive waves and methods of their construction}\label{sec:igw}

From the physical point of view, an impulsive gravitational wave
most naturally can  be understood as a limit of a suitable family of
gravitational waves with sandwich profiles of ever ``shorter
duration'' $\varepsilon$ which simultaneously become
``stronger'' as ${\varepsilon^{-1}}$.
Mathematically, this amounts to a distributional limit in which a sequence of
sandwich profiles converges to the profile $\delta(u)$, the
Dirac function. An impulsive gravitational wave is thus localized on a
\emph{single wave-front} ${u=0}$, which is a null
hypersurface. Across ${u=0}$ the first derivative of the metric with respect to
$u$ is discontinuous, introducing a Dirac delta in the curvature tensor representing the gravitational impulse.

Interestingly, there exist several alternative methods of construction of such
exact nonexpanding solutions to Einstein's vacuum field equations. They will now be
summarized and compared, together with the appropriate references to
original works.

\subsection{The ``cut and paste'' method}

Let us start with an elegant geometrical method for constructing 
impulsive plane gravitational waves in Minkowski background presented by
Penrose in now classic works \cite{Pen68a,Pen68b,Pen72}.
His ``cut and paste'' approach\footnote{Penrose's approach is in some respect similar
to that of Israel \cite{Israel66, BarIsr91} and closely related to
the Dray and 't~Hooft method of ``shift function'' \cite{DaT}.}
is based on the removal of the null hypersurface ${\cal N}$ given
by $\U=0$ from the spacetime 
 \begin{equation}
\d s_0^2= \frac{2\,\d\eta\,\d\bar\eta
-2\,\d\U\,\d\V}{[\,1+{\frac{1}{6}}\Lambda(\eta\bar\eta-\U\V)\,]^2}\,,
 \label{conf*}
 \end{equation}
(with ${\Lambda=0}$), and re-attaching the ``halves'' ${{\cal M}^-(\U<0)}$ and ${{\cal
M}^+(\U>0)}$ by making an identification with a ``warp''
in the coordinate $\V$ such that (see Figure~\ref{planecut})
\begin{equation}
\Big[\,\eta,\,\bar\eta,\,\V,\,\U=0_-\,\Big]_{_{{\cal M}^-}}\equiv
\Big[\,\eta,\,\bar\eta,\,\V-H(\eta,\bar\eta),\,\U=0_+\,\Big]_{_{{\cal M}^+}}\,,
\label{juncnon}
\end{equation}
where $H(\eta,\bar\eta)$ is an {\it arbitrary} real-valued function of $\eta$
and $\bar\eta$.
It was shown in \cite{Pen72} that
impulsive components are introduced into the curvature tensor
proportional to $\delta(\U)$ representing gravitational (plus
possibly null-matter) impulsive waves.
\bigskip \\

\begin{minipage}[b]{.44\textwidth}
\centering
\includegraphics[width=0.75\textwidth]{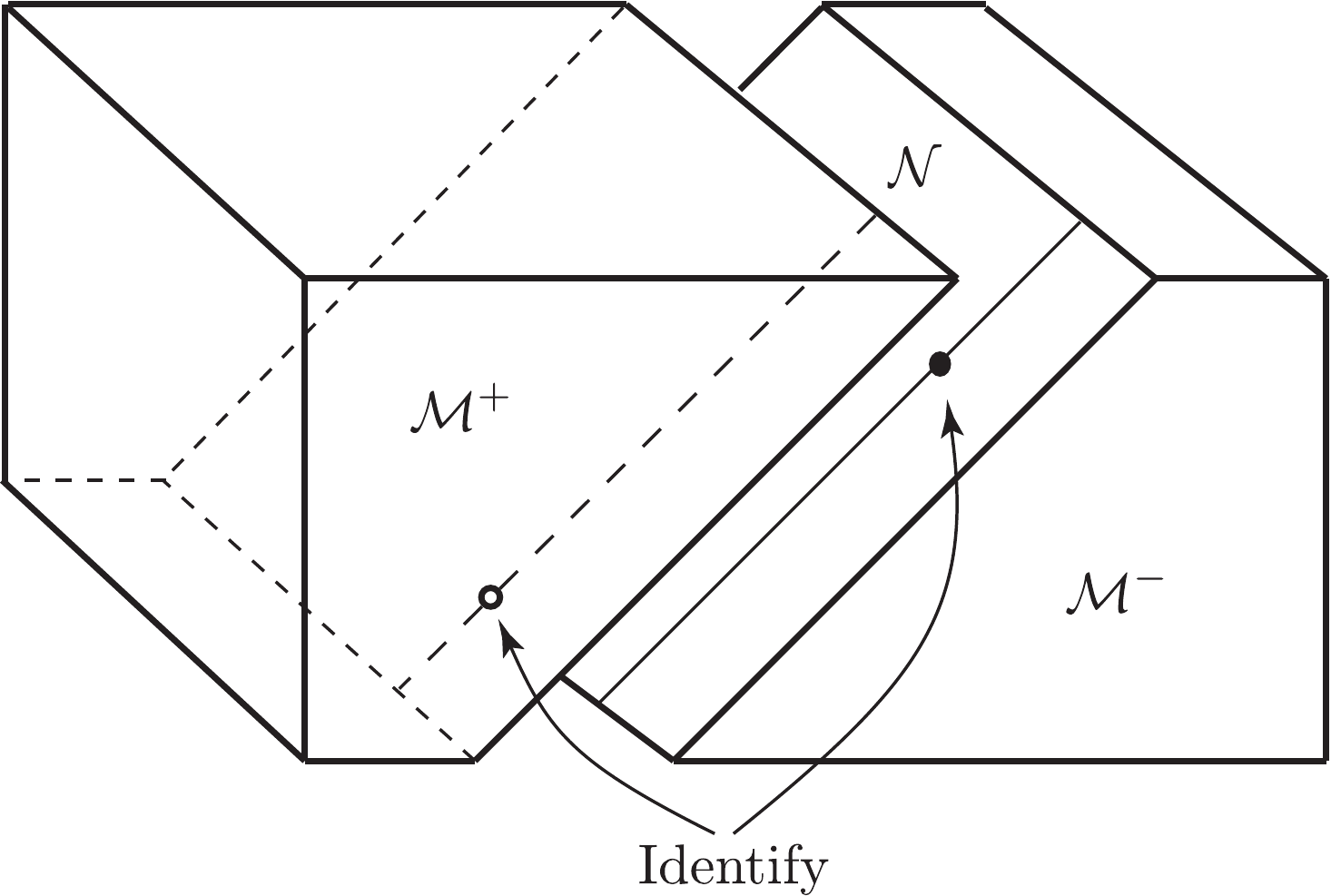}
\label{planecut}
\end{minipage}
\hfill
\begin{minipage}[b]{.56\textwidth}
 { \small Figure~\ref{planecut}: Minkowski space is cut into two parts ${\cal
M}^-$ and ${\cal
M}^+$ along the null hyperplane~${\cal
N}$. The parts are then re-attached with a ``warp'' in which points
are shunted along the null generators of the
cut and then identified. The same construction in de Sitter or anti-de Sitter
space generates impulsive gravitational waves
in these backgrounds.}\\
\end{minipage}
\smallskip

In \cite{Pen72} Penrose considered only a Minkowski background,
in which case the impulsive surface ${\,\U=0}$ is obviously a plane, and impulsive
{\it pp\,}-waves are obtained. In \cite{[B7]} it was demonstrated
that exactly the same junction conditions (\ref{juncnon}) applied to a
general background spacetime (\ref{conf*}) of constant curvature introduce impulsive waves
also in the de Sitter (${\Lambda>0}$) or anti-de~Sitter ($\Lambda<0$)
universes. However, the geometries of these impulses are different since
the null hypersurface ${\,\U=0}$, along which the spacetime is cut and
pasted, is described by the 2-metric
${\d\sigma^2= 2\,(\,1+\frac{1}{6}\Lambda\eta\bar\eta)^{-2}\,\d\eta\,\d\bar\eta}$.
This is a 2-dimensional space of constant Gaussian curvature ${K=\frac{1}{3}\Lambda}$, which
is a plane for ${\Lambda=0}$. In the ${\Lambda\ne0}$ cases it is either
a {\it sphere} (${\Lambda>0}$) or a {\it hyperboloid} (${\Lambda<0}$). These geometries
have been described in detail in~\cite{[B5]} using various
coordinate representations, and it was explicitly demonstrated that the wave
surfaces are nonexpanding.

\subsection{The continuous form of the metric}

While the ``cut and paste'' method describes the identification of points
on both sides of the impulse it does not provide explicit metric forms of the complete
spacetimes. Thus the next step is to find a suitable coordinate system in which the metric
is continuous as a function of $\U$.

Starting again with the metric (\ref{conf*}) we perform the transformation
\begin{equation}\label{ro:trsf}
 \U=U\,,\quad \V=V+H+UH_{,Z}H_{,\bar Z}\,,\quad \eta=Z+UH_{,\bar Z}\,,
\end{equation}
where $H(Z,\bar Z)$ is an arbitrary real-valued function.
This yields the metric
 \begin{equation}
\d s_0^2= \frac{2\,|\d Z+U(H_{,Z\bar Z}\d Z+H_{,\bar Z\bar Z}\d\bar Z)|^2-2\,\d U\d V}
{[\,1+\frac{1}{6}\Lambda(Z\bar Z-UV-UG)\,]^2}\,,
 \label{deS2}
 \end{equation}
 where ${G(Z,\bar Z)\equiv H-ZH_{,Z}-\bar ZH_{,\bar Z}}$.
We now consider (\ref{deS2}) for ${U>0}$ and combine this with the line element
(\ref{conf*}) in which we set ${\U=U}$, ${\V=V}$, ${\eta=Z}$ for ${U<0}$. The resulting
metric can be written as
 \begin{equation}
\d s^2= \frac{2\,|\d Z+\Up(H_{,Z\bar Z}\d Z+H_{,\bar Z\bar Z}\d\bar Z)|^2-2\,\d U\d V}
{[\,1+\frac{1}{6}\Lambda(Z\bar Z-UV-\Up G)\,]^2}\,,
 \label{conti}
 \end{equation}
 where
\begin{align}
\Up\equiv\Up(U) &=
    \begin{cases}
     0 & \text{if } U \leq 0\,, \\
     U & \text{if } U \geq 0
    \end{cases}
\end{align}
is the \emph{kink-function}. Since the kink function is Lipschitz continuous the
metric (\ref{conti}) is locally Lipschitz in the variable $U$. Thus, apart from
possible singularities of the function $H$ and its derivatives (which indeed occur in physically
realistic models, see e.g.\ section~\ref{sec:boost}, below), the spacetime
is locally Lipschitz. Observe that any locally Lipschitz metric $g$ (denoted by ${g\in\Con^{0,1}}$) 
possesses a locally bounded connection and so the curvature is a distribution. Also
we are well within the ``maximal'' distributional curvature framework as identified
by Geroch and Traschen \cite{GT:87}. For locally Lipschitz metrics
there is no bound on the curvature (in $L^\infty$). Indeed, the discontinuity in
the derivatives of the
metric introduces  impulsive components in the Weyl and curvature tensors \cite{[B7]}, namely
${\,\Psi_4 = (1+\frac{1}{6}\Lambda Z\bar Z)^2 H_{,ZZ} \,\delta(U)\,}$ and
${\,\Phi_{22} = [\, (1+\frac{1}{6}\Lambda Z\bar  Z )^3
 ((1+\frac{1}{6}\Lambda Z\bar Z)^{-1}H)_{,Z\bar Z}
+\frac{1}{3}\Lambda \,H\, ]\, \delta(U)\,}$, in a natural tetrad.
The metric (\ref{conti}) thus explicitly describes  impulsive
waves in de~Sitter, anti-de~Sitter or Minkowski backgrounds. For ${\Lambda=0}$,
the conformal factor is $1$, and the line element (\ref{conti}) reduces to the Rosen form of impulsive
{\it pp\,}-waves \cite{Pen72,[B2],Steinb}.

The transformations relating (\ref{conf*}) and (\ref{conti}) separately for ${U<0}$ and
${U>0}$ can be written  in a combined way using the \emph{Heaviside function}
${\Theta=\Theta(U)}$ as
\begin{equation} \label{trans}
 \U=U\,,\quad
 \V=V+\Theta\,H+\Up\,H_{,Z}H_{,\bar Z}\,, \quad
 \eta=Z+\Up\,H_{,\bar Z}\,,
\end{equation}
which is discontinuous in the coordinate $\V$ on ${\{\U=0\}}$. From (\ref{trans})
we, in particular, obtain the Penrose junction condition (\ref{juncnon})
for reattaching ${\cal M}^-$ and ${\cal M}^+$ with a warp.
Thus, the above procedure is indeed an {\it explicit} Penrose's ``cut and paste''
construction of all nonexpanding impulsive gravitational waves.

\subsection{The distributional form of the metric}

The most intuitive way of constructing impulsive waves is the \emph{distributional limit} of suitable
families of gravitational waves with smooth sandwich profiles alluded to above.
For vacuum {\it pp\,}-waves, such a procedure
was considered  in  \cite{Pen68a, Pen68b,Rindler} and later
elsewhere (e.g. \cite{[A3]}). In this simplest case  one obtains the metric
 \begin{equation}
\d s^2=2\,\d\xi\,\d\bar\xi - 2\,\d u\,\d v+\H(\xi,\bar\xi)\,\delta(u)\, \d u^2  \,,\label{imp-pp}
\end{equation}
which is  the well-know Brinkmann form of impulsive {\it pp}-waves in  Minkowski background
\cite{Pen68a, Pen68b,Pen72}.

More general impulsive waves within the Kundt class (see \cite{Kundt},
\cite[Ch.\ 31]{KSMH}, \cite[Ch.\ 18]{GP:09}) of nonexpanding spacetimes
with ${\Lambda\ne0}$ can be obtained similarly. It was demonstrated in \cite{[B1]} that
{\it all} nonexpanding impulses in Minkowski or (anti-)de~Sitter universes can
be constructed from the general class of type~N Kundt solutions \cite{ORR, [A1]},
upon considering the  distributional limit $\H(\xi,\bar\xi)\,\delta(u)$
of the structural function, i.e.,
\begin{equation}
\d s^2=\frac{2}{P^2}\,\d\xi\,\d\bar\xi - 2\, \frac{Q^2}{P^2}\,\d u\,\d v
+ \bigg[2k\frac{Q^2}{P^2}\,v^2 - \frac{(Q^2)_{,u}}{P^2}\,v + \frac{Q}{P} \,\H(\xi,\bar\xi)\,\delta(u)\bigg] \d u^2\
,\label{KNimp}
\end{equation}
where ${\,P=1+ \frac{1}{6}\Lambda \xi\bar\xi}$,
${\,Q=(1-\frac{1}{6}\Lambda\xi\bar\xi)\,a +\bar b\,\xi+b\,\bar\xi}$,
and ${\,k=\frac{1}{6}\Lambda a^2+b\bar b\,}$. The metric (\ref{KNimp}) contains
a single $\delta(u)$ and obviously generalizes the Brinkmann form of impulsive
{\it pp\,}-waves (\ref{imp-pp}) to which it reduces when ${P=1=Q}$ for
${\Lambda=0}$.

Of course, a distributional term in the metric leads us out of the
Geroch--Traschen class \cite{GT:87} of metrics, for which the metric is of
regularity $W^{2,1}_{\mbox{\small loc}}\cap L^\infty_{\mbox{\small loc}}$, which
guarantees the curvature to exist in distributions. However, due to its simple geometrical structure
the metric (\ref{KNimp})  nevertheless allows to calculate the curvature as a distribution.

For a {\it generic} function $\H(\xi,\bar\xi,u)$ there
exist distinct canonical subclasses of type~N Kundt solutions characterized
by specific choices of the parameters $a$ and $b$, see \cite{ORR, [A1], GP:09}.
Surprisingly, it was proven in \cite{[B1]} that {\it impulsive}
limits $\H(\xi,\bar\xi)\,\delta(u)$ of these  subclasses for a given $\Lambda$ become
(locally) {\it equivalent}. For example, in the case of ${\Lambda=0}$ there are two
subclasses, namely the  {\it pp\,}-waves and the Kundt waves $KN$. However, the transformation
\begin{equation}
   \U   = (\xi+\bar\xi)(1+uv)u\,, \qquad
   \V   = (\xi+\bar\xi)v-1      \,, \qquad
   \eta = \xi +(\xi+\bar\xi)uv \,, \label{E3.1}
\end{equation}
converts the impulsive ${KN}$ metric (\ref{KNimp}) with ${a=0}$, ${b=1}$ to
the impulsive {\it pp\,}-wave metric with ${a=1}$, ${b=0}$.
The only non-trivial impulsive gravitational waves of the form (\ref{KNimp}) in Minkowski space are thus the impulsive
{\it pp\,}-waves (\ref{imp-pp}). Similar results hold also for ${\Lambda>0}$ and ${\Lambda<0}$. Generically, there are
three distinct subclasses of nonexpanding waves, namely  ${KN(\Lambda)I}$ given by ${a=0}$, ${b=1}$,
${KN(\Lambda^-)II}$ given by ${a=1}$, ${b=0}$, ${\Lambda<0}$, and generalized Siklos waves \cite{Sik, [A6]}
${KN(\Lambda^-)III}$
for which ${k=0}$ and ${\Lambda<0}$, see \cite{GP:09}.
In all these cases it was shown in \cite{[B1]} that although these canonical
subclasses  are different for extended profiles, they are  equivalent for
impulsive profiles. In this way, by considering the above distributional limit (\ref{KNimp})
of the class we obtain an explicit form of \emph{all} solutions representing
nonexpanding impulses.

Interestingly, there exists yet another metric form of representing this complete
family of impulsive solutions. It is obtained from the continuous form of the
impulsive wave metric (\ref{conti}) by applying the transformation (\ref{trans})
not separately for ${U<0}$ and ${U>0}$ but (formally) for all values of $U$ including
on the impulse. Explicitly, if we keep the terms arising from the derivatives of
$\Theta$, this transformation relates (\ref{conti}) to
 \begin{equation}
\d s^2= \frac{2\,\d\eta\,\d\bar\eta-2\,\d \U\,\d \V +2H(\eta,\bar\eta)\,\delta(\U)\,\d \U^2}
{[\,1+\frac{1}{6}\Lambda(\eta\bar\eta-\U\V)\,]^2}\,.
 \label{confppimp}
 \end{equation}
Observing also from \eqref{trans} that $\eta=Z$ if $U=0=\U$, the function ${H(Z,\bar Z)}$
of \eqref{conti} agrees with ${H(\eta,\bar\eta)}$ of \eqref{confppimp} on the wave surface.
Again, in Minkowski background this is just the Brinkmann form of a general impulsive
{\it pp\,}-wave (\ref{imp-pp}).\footnote{In fact, the metric
(\ref{confppimp}) is conformal to (\ref{imp-pp}). Recall in this context that
Siklos \cite{Sik} proved that Einstein spaces conformal to {\it pp\,}-waves
only occur when ${\Lambda<0}$. The \emph{impulsive} case shows that this result does not hold
in low regularity.}

The explicit transformation relating (\ref{KNimp}) to
(\ref{confppimp}) for the $KN(\Lambda)I$ subclass is
\[
\U   = \frac{(\xi+\bar\xi)(1+uv)u}
  {1-\frac{1}{6}\Lambda(\xi+\bar\xi)(1+uv)u}\,,\  
\V  = \frac{(\xi+\bar\xi)v+\frac{1}{6}\Lambda\xi\bar\xi}
  {1-\frac{1}{6}\Lambda(\xi+\bar\xi)(1+uv)u} -1\,,\  
\eta  = \frac{\xi +(\xi+\bar\xi)uv}
  {1-\frac{1}{6}\Lambda(\xi+\bar\xi)(1+uv)u}\,, 
\]
which reduces to (\ref{E3.1}) in the case ${\Lambda=0}$. Similar
transformations exist also for the subclasses ${KN(\Lambda^-)II}$ and
${KN(\Lambda^-)III}$, see \cite{[B1]}. Therefore, the full family of
impulsive limits (\ref{KNimp})
of  nonexpanding sandwich waves of the Kundt class
is indeed equivalent to the distributional form of the solutions
 (\ref{confppimp}), and consequently to the continuous metric (\ref{conti})
 obtained by the ``cut and paste'' method.

Of course, the discontinuity in the complete transformation (\ref{trans})
is mathematically delicate.
However, in the special case of impulsive {\it pp\,}-waves it was rigorously
analyzed using a general regularization procedure in \cite{Steinb}.
Indeed it was shown within the geometric theory of nonlinear generalized
functions \cite{GKOS:01} (based on Colombeau algebras \cite{C:85})
that (\ref{trans}) is a (generalized) coordinate transformation,
a result which puts the formal (``physical'') equivalence of  both
forms of impulsive spacetimes on a solid ground.\footnote{Interestingly, these studies
have triggered a corresponding line of research in generalized functions
\cite{EG:11,EG:13}.}

\subsection{Boosting static sources}\label{sec:boost}

As demonstrated in 1971 by Aichelburg and Sexl in a  classic paper
\cite{AS}, a specific impulsive gravitational {\it pp\,}-wave solution
(in distributional form) can be obtained by boosting the Schwarzschild black hole to the speed of light,
while its mass is scaled to zero in an appropriate way. Such a solution represents an
axially-symmetric impulsive gravitational wave in Minkowski space
generated by a single {\it null monopole} particle moving along the axis.
Note that the continuous coordinate system for the Aichelburg--Sexl solution
was found by D'Eath \cite{DE78} and used for investigation of ultrarelativistic black-hole encounters.

Using a similar approach, numbers of other specific impulsive
waves in flat space have been obtained by boosting more general black hole
spacetimes \cite{FerPen90, LouSan92, BalNach95, BalNach96, Yoshino05}, multipole
sources \cite{[B4]} or black rings \cite{OrtKrtPod05,OrtPodKrt05}. This method
has been generalized to the ${\Lambda\not=0}$-cases by Hotta and
Tanaka~\cite{HotTan93}, who
boosted the Schwarzschild--de~Sitter solution to obtain a nonexpanding spherical
impulsive gravitational wave generated by a pair of null monopole particles in
the de~Sitter background. They also described an analogous solution in the
anti-de~Sitter universe. Their main ``trick'' was to consider the boost in the 5-dimensional
representation of the (anti-)de~Sitter spacetime (see also section
\ref{subsec:emb}, below), where the boost can explicitly (and consistently) be
performed.

Details on  boosting monopole particles
to the speed of light in the (anti-)de~Sitter universe, the geometry of
the nonexpanding wave surfaces, and discussion of
various useful coordinates can be found in~\cite{[B5]}.
It was also shown that although the impulsive
wave surface is nonexpanding, for ${\Lambda>0}$ this coincides with
the horizon of the closed de~Sitter universe. The background space
contracts to a minimum size and then re-expands in such a way that
the impulse in fact propagates with the speed
of light from the ``north pole'' of the universe across the
equator to its ``south pole''.

There are also particular impulsive waves generated by
{\it null multipole} particles obtained by  boosting static multipole sources \cite{[B4]}.
Such solutions with ${\Lambda=0}$ can be written in the form (\ref{imp-pp}) with
\begin{equation}\label{Hsing}
 \H= \textstyle{-b_0\log\rho +\sum_{m=1}^\infty b_m\,\rho^{-m}\cos[m(\phi-\phi_m)]}\,,
\end{equation}
where ${\xi=\frac{1}{\sqrt2}\,\rho\, e^{i\phi}}$  and $b_m$, $\phi_m$ are constants.
The term given by ${b_0\log\rho}$ represents the Aichelburg--Sexl solution \cite{AS} for a single null monopole particle. The
terms with ${m\ge1}$ correspond to the multipole components of an impulsive {\it pp\,}-wave generated by a source of an
arbitrary multipole structure \cite{[B3]}. Indeed, the field equations relate them  to a source localized  at ${\rho=0}$ on
the impulsive wavefront ${u=0}$, which is described by $T_{uu}=J(\rho,\phi)\,\delta(u)$ with
 \begin{equation}
J(\rho,\phi)={\textstyle\frac{1}{4} b_0\,\delta(\rho) + \sum_{m=1}^\infty
\frac{1}{4}b_m  \frac{(-1)^m}{(m-1)!}\,\delta^{(m)}(\rho)\cos[m(\phi-\phi_m)]}\,.
 \end{equation}

Observe that any function $\H$ of the form \eqref{Hsing} is singular on
the axis $\rho=0$ leading to curvature singularities in the spacetime at
${\rho=0}$, ${u=0}$.

Interestingly, as demonstrated in \cite{[B6]}, there are analogous
impulsive solutions also
in the case ${\Lambda\not=0}$. For their description it is, however, more
convenient to use the formalism based on embedding
the (anti-)de Sitter universe into the 5-dimensional Minkowski space, as we
will detail next.

\subsection{Embedding to 5 dimensions}\label{subsec:emb}

The full class of nonexpanding impulsive waves in spaces of constant
curvature with ${\Lambda\ne0}$ can
be obtained in a 5-dimensional formalism as metrics
\begin{equation}
\d s^2 =  \d{Z_2}^2 +\d{Z_3}^2+\epsilon\,\d{Z_4}^2- 2\d\tilde U\d\tilde V 
  +\H(Z_2,Z_3,Z_4)\,\delta(\tilde U)\,\d\tilde U^2\,,
\label{general}
\end{equation}
with the constraint
${Z_2}^2+{Z_3}^2+\epsilon{Z_4}^2-{2\tilde U\tilde V=\epsilon a^2}$, 
where  ${\tilde{U}=\frac{1}{\sqrt2}(Z_0+Z_1)}$, ${\tilde{V}=\frac{1}{\sqrt2}(Z_0-Z_1)}$, 
${a=\sqrt{3/|\Lambda|}}$ and ${\epsilon =
\hbox{sign}\,\Lambda}$.
As shown in \cite{[B6]} this metric represents impulsive waves propagating in the
(anti-)de~Sitter universe with the impulse located on the null hypersurface 
${\tilde U=0}$, i.e.,
 \begin{equation}
 {Z_2}^2+{Z_3}^2+\epsilon{Z_4}^2=\epsilon a^2\,,\label{surface}
 \end{equation}
which is a {\it nonexpanding} 2-sphere in the de~Sitter universe and a
hyperboloidal 2-surface in the anti-de~Sitter universe, respectively. Various
4-dimensional coordinate parametrizations of
(\ref{general}) can be considered. For example,
\begin{equation}
\tilde{U} = \frac{\U}{\Omega}\,, \quad   \tilde{V} = \frac{\V}{\Omega} \,, \quad 
 Z_2+iZ_3= \frac{x}{\Omega}+i \frac{y}{\Omega} = \frac{\sqrt2\,\eta}{\Omega}\,,  \quad
 Z_4 = a \left(\frac{2}{\Omega}-1\right), \label{Zcoord}
 \end{equation}
where ${\,\Omega=1+{\textstyle\frac{1}{6}}\Lambda(\eta\bar\eta-\U\V)=1+{\textstyle\frac{1}{12}}\Lambda(x^2+y^2-2\U\V)}$,
brings the metric to the previous form  (\ref{confppimp}) with the function
\begin{equation}
 \H=2 H/{\textstyle(1+\frac{1}{6}\Lambda\eta\bar\eta)}\,.
\end{equation}
Other natural coordinates which parametrize (\ref{general}) have been discussed
in \cite{[B5]}.

The metric (\ref{general}) may describe
impulsive gravitational waves and/or impulses of null matter. Purely gravitational waves occur when
the   vacuum field equation
 \begin{equation}
(\Delta+{\textstyle\frac{2}{3}}\Lambda)\,\H=0 \label{vacuum}
 \end{equation}
is satisfied \cite{HorItz99, Sfet, [B7]}, where
${\Delta\equiv \frac{1}{3}\Lambda\{\partial_z[(1-z^2)\partial_z]+(1-z^2)^{-1} \partial_\phi\partial_\phi)\}}$
is the Laplacian on the impulsive surface (\ref{surface}), parametrized by
$Z_2=a\sqrt{\epsilon(1-z^2)}\,\cos\phi$, $Z_3=a\sqrt{\epsilon(1-z^2)}\,\sin\phi$, and
$Z_4=a\,z$. It was demonstrated in  \cite{[B6]} that nontrivial
solutions of  (\ref{vacuum}) can be written as
 \begin{equation}
\H(z,\phi)= \textstyle{ b_0\,Q_1(z)+\sum_{m=1}^\infty b_m\,Q^m_1(z)\cos[m(\phi-\phi_m)]}\,, \label{E4.5}
\end{equation}
where  $Q^m_1(z)$ are associated Legendre functions of the second kind generated
by the relation
${Q^m_1(z)=(-\epsilon)^m|1-z^2|^{m/2}\frac{\d^m}{\d z^m}Q_1(z)}$.
The first term for ${m=0}$, i.e., ${Q_1(z)=\frac{z}{2}\log\left|\frac{1+z}{1-z}\right|-1}$,
represents the simplest axisymmetric Hotta--Tanaka solution \cite{HotTan93}.
The components with ${m\ge1}$ describe nonexpanding impulsive gravitational
waves in (anti-)de~Sitter universe generated by null point sources with an
$m$-pole structure, localized on  the wave-front at the singularities
${z=\pm1}$.

\subsection{Summary of the construction methods}

To end this review we collect the various methods of constructing nonexpanding impulsive
waves in spaces of constant curvature and the corresponding references in the following table:

\begin{center}
\begin{tabular}{|l||c|c|}
\hline
   method of construction & ${\Lambda=0}$   & ${\Lambda\not=0}$ \\
\hline
\hline
  ``cut and paste'' & \cite{Pen68a}, \cite{Pen68b}, \cite{Pen72}, \cite{DaT} & \cite{[B7]}, \cite{Sfet} \\
\hline
  continuous coordinates   & \cite{Pen72}, \cite{DE78}, \cite{[B2]}, \cite{Steinb} & \cite{[B7]}, \cite{[B1]} \\
\hline
 limits of sandwich waves   &  \cite{Pen68a}, \cite{Pen68b}, \cite{Rindler}, \cite{[A3]},  \cite{Steina}, \cite{KS:99}, \cite{[B1]}  & \cite{[B1]}, \cite{[B6]} \\
\hline
  boosts &  \cite{AS}, \cite{FerPen90}--\cite{OrtPodKrt05}, \cite{[B3]} & \cite{HotTan93}, \cite{[B5]}, \cite{[B4]} \\
\hline
  embedding  & --- &  \cite{HotTan93}, \cite{[B5]},  \cite{[B6]}, \cite{HorItz99},  \cite{[B7]} \\
\hline
\end{tabular}
\end{center}

For more details see, e.g., \cite{Pod2002b}, \cite{BarHog2003} and \cite[Ch.\ 20]{GP:09}.

\section{Geodesics in nonexpanding impulsive waves}\label{sec:geo}

In this section we shift our focus on the main theme of this work, i.e., the
analysis of the geodesic equation in spacetimes with nonexpanding impulsive waves. While we
will mainly be concerned with the continuous form of the metric (\ref{conti})
we start by reviewing some results obtained using the distributional form of
the metric (\ref{confppimp}).

\subsection{Distributional form}\label{sec:3.1}

Geodesics in Minkowski space with impulsive {\it pp\,}-waves
were discussed in many works, e.g.\ in  \cite{DaT, Sfet, DT, Bal1997,[C2]}, all
deriving that they are refracted straight lines with a jump in the
$\mathcal{V}$-direction. However, the corresponding geodesic (and
also the geodesic deviation) equations in standard coordinates
(\ref{imp-pp}) cannot be formulated consistently in distributions,
since they contain ill-defined products. These equations have been rigorously
analyzed in
\cite{Steina,KS:99} using the geometric theory of nonlinear generalized
functions~\cite{GKOS:01}.
In particular, existence and uniqueness results have been obtained in a space
of generalized functions and the geodesics have been shown again to be broken straight lines. 
The benefit of the rigorous approach is the
following: There it is \emph{proven} that the geodesics really cross the
impulsive wave rather than being reflected or trapped, a possibility which is ruled out a
priori by the approaches using multiplications rules and other tricks from the
gray area of (linear) distribution theory. As a consequence, in the rigorous
approach \emph{geodesic completeness} of impulsive {\it pp}-waves is proven---a
result which has been recently generalized to models allowing for a non-flat
wave surface \cite{SS12,SS14}.

In \cite{PO:01} the geodesics in nonexpanding impulsive waves in all constant
curvature backgrounds with any $\Lambda$ have been studied. They have been derived using the
embedding of (anti-)de~Sitter spacetime into $5$-dimensional Minkowski space as
detailed in section \ref{subsec:emb}. The advantage of this approach
lies in the fact that it yields a system of differential equations which
is distributionally accessible at all, if not rigorously. In particular, there
is no square of   $\delta$, contrary to other ``direct'' approaches, such as
those of \cite{Sfet} which used the coordinates introduced in \cite{DaT}.
The general results of \cite{PO:01} in the special case
${\Lambda=0}$ reduce to those {\it rigorously} derived in \cite{KS:99}.
Nevertheless, a desirable nonlinear distributional analysis of the geodesic
equation in the ${\Lambda\not=0}$-cases is still subject to ongoing research.

\subsection{Continuous form}\label{sec:3.2}

Recently the continuous form of the metric \eqref{conti} has been employed in \cite{LSS:13} to
derive the geodesics in impulsive {\it pp\,}-waves. Here the geodesic
equation, which has a discontinuous right hand side, has been \emph{uniquely
solved in the sense of Carath\'eodory}. In this way it has, in particular, been
shown that the spacetimes are geodesically complete with the geodesics being
\emph{continuously differentiable} curves. This justifies the
$\mathcal{C}^1$-matching of the geodesics of the background to obtain the
geodesics of the entire spacetime. In fact such an approach has been used in
\cite{[C2],bis-proc}, as well as in \cite{[B7],PO:01} for nonvanishing $\Lambda$,
and in \cite{PS03,PS10} for expanding impulsive waves.

The analysis of \cite{LSS:13} is based on the fact that for
{\it pp\,}-waves the coordinate $U$ can be used as a parameter along the
geodesics. However, this is no longer possible if ${\Lambda\ne0}$, where the
geodesic equations take the form of an autonomous system of ODEs with
discontinuous right hand side. In this case Carath\'eodory's concept
provides no advantage over the classical theory and is not applicable to the
equations at hand.

On the other hand it was also recently shown in \cite{S:14} that
the geodesic equations for any locally Lipschitz continuous semi-Riemannian
metric possess solutions \emph{in the sense of Filippov} and that these geodesics in
addition are continuously differentiable. Filippov's solution concept
\cite{F:88} is a general and nowadays widely applied approach (e.g.\ in nonsmooth
mechanics, see \cite{C:08}) and, for the convenience of the reader, we have
collected its basics used in this work in appendix \hyperref[app:a]{A}. Since we
will make use of this result in the following, we give its precise formulation:
\begin{thm}(\cite[Theorem 2]{S:14})\label{thm-ex}
Let $(M,g)$ be a smooth manifold with a $\Con^{0,1}$-semi-Riemannian metric $g$. Then there
exist Filippov solutions of the geodesic equations which are $\Con^1$-curves.
\end{thm}

However, in general we cannot expect the geodesic equation in locally
Lipschitz spacetimes to be \emph{uniquely} solvable. The threshold for unique
solvability of the geodesic equations is the regularity class
$\mathcal{C}^{1,1}$, i.e., the first derivatives of the metric being locally
Lipschitz. In fact, in this class classical ODE-theory provides unique
solvability of the geodesic equation with the geodesics being~$\mathcal{C}^2$.
Moreover, as has been recently shown \cite{KSS14,M13}, the exponential map
retains maximal regularity.
On the other hand there exist metrics in any H\"older class
$\mathcal{C}^{1,\alpha}$, with ${\alpha<1}$, for which the initial value problem
for the geodesic equation fails to be uniquely solvable. Here we
recall the following classical example due to \cite{Hartman:50} for a $\Con^1$
(hence, in particular, locally Lipschitz) Riemannian metric.  Consider the line
element $\dd s^2= {h(y)(\dd x^2+\dd y^2)}$ on $\R^2$ with the function $h(y)\equiv 1 +
y^{\frac{4}{3}}$. The corresponding geodesic equations then
reduce to
\begin{equation*}
 2 \frac{\dd^2 y}{\dd x^2} = \left(1 + \Big(\frac{\dd y}{\dd x}\Big)^2\right)\frac{\dd}{\dd y}\log(h(y))\,.
\end{equation*}
Thus $y(x)=0$ and $y(x)=x^3$ are two distinct solutions starting at $0$ with
initial velocity $0$.

For our purpose it is worth to observe that these geodesics are
classical solutions ($\Con^2$ and they satisfy the differential equation
everywhere) so they are also Filippov solutions.
Consequently in this case we have non-unique Filippov solutions.

However, in the case we are interested in, that is the continuous
form of the metric for nonexpanding impulsive waves in all constant curvature
backgrounds (\ref{conti}), the metric \emph{in addition} to being locally Lipschitz is
also \emph{smooth off a null hypersurface}. In particular, it is piecewise smooth and
in such a case uniqueness of the geodesics can indeed be established (as we will see next),
thereby justifying the $\mathcal{C}^1$-matching procedure.

\subsection{Unique \texorpdfstring{$\mathcal{C}^1$}{C1}-geodesics for a class of locally Lipschitz metrics}\label{sec:3.3}

To formulate the result announced above we need some preparations. We
consider a spacetime $(M,g)$ with \emph{locally Lipschitz continuous}
metric $g$ and global null coordinates $U,V$ such that $M$ is separated into
two parts by a \emph{totally geodesic} null hypersurface $N=\{U=0\}$. By $N$ being
totally geodesic we mean that every geodesic (in the sense of Filippov) starting in $N$ and being initially
tangential to $N$ stays (initially) in $N$.

\begin{rem}
Observe that the notion of a totally geodesic (null) hypersurface $N$ of a spacetime $(M,g)$ with $g\in\Con^{0,1}$
is somewhat subtle. Indeed, if geodesics are not unique, then from the
classically equivalent characterizations
\begin{enumerate}
 \item [(1)] the second fundamental form of $N$ vanishes (in $L_\loc^\infty$),
 \item [(2)] every geodesic in $N$ is a geodesic in $M$,
 \item [(3)] every geodesic starting in $N$ tangential to $N$ stays (initially) in $N$,
\end{enumerate}
condition (3) (i.e., our definition) becomes stronger while it implies the still equivalent conditions (1) and (2).
Note, however, that in the spacetimes we are dealing with we prove uniqueness of
geodesics and so (1), (2), and (3)
become again equivalent.
\end{rem}

We want to rewrite the geodesic
equation in \emph{first order form}. To this end we denote by $X,Y$ some local spatial
coordinates and introduce
\begin{equation}\label{notation Xa}
\bW \equiv (U,\hat U,V,\hat V,X,\hat X,Y,\hat Y) \equiv (U,\dot U\!, V,
\dot{V}\!, X, \dot X\!, Y, \dot Y\!)\,.
\end{equation}
We assume that the geodesic equation for $(M,g)$, as a first order system, has
the form
\begin{equation}\label{eq-de}
\dot W^a = \frac{E^a(\bW)}{P^a(U,V,X,Y)} \Theta(U) +
\frac{F^a(\bW)}{Q^a(U,V,X,Y)} \equiv
T^a(\bW)\,,
\end{equation}
for $a=1,\ldots,8$ labeling the components $W^a$ of $\bW$ in \eqref{notation
Xa}, and analogously for $\bE$, $\bF$, $\bP$, $\bQ$, $\bT$. Here we further
assume that $\bE$, $\bF$ are \emph{smooth}, and that
$\bP$, $\bQ$ are \emph{smooth} in ${V,X,Y}$ but are \emph{polynomial in
$\Up$} (of degree at
most five), locally bounded away from zero, hence $\bT\in
L_\loc^\infty$.

Note that
$\Up$ is Lipschitz continuous and $\Theta$ is discontinuous at $U=0$, so by
restricting ourselves to the form \eqref{eq-de} of the geodesic equation we allow only for
non-smoothness in $U$ at $U=0$ given by a jump
($\Theta$) and possibly by (higher order) kinks ($U_+$, $U_+^2$, etc.).
Now we may state and prove the following:

\begin{thm}\label{mainthm}
Let $(M,g)$ be a spacetime with a $\Con^{0,1}$-metric $g$ as above and assume the
geodesic equation to be of the form \eqref{eq-de}.
Then given initial data with ${U\ne0}$ the geodesic equation possesses unique
$\Con^1$-solutions in the sense of Filippov.
\end{thm}

\begin{pr} Existence follows from Theorem \ref{thm-ex}, since the metric is
locally
Lipschitz continuous by assumption. Therefore it only remains to prove
\emph{uniqueness} and we aim at applying Corollary \ref{cor-app-uni} of the appendix.

Rewriting the geodesic equations as first order system as above and setting
${D^-\equiv \{\bW: U < 0\}}$, ${D^+\equiv\{\bW: U > 0\}}$, ${N\equiv \partial
D^-
= \partial
D^+ = \{\bW:U=0\}}$, we obviously get that $\bT$ is smooth except
on $N$, were it is discontinuous,
hence it is piecewise continuous. Moreover $\bT$ satisfies
\begin{align}
 (T^a)^- &= \Big( \frac{F^a}{Q^a}\Big)\Big|_{\bar D^-} \in \Cinf(\bar
D^-)\,, \\
 (T^{a})^+ &= \Big(\frac{E^a}{P^a} + \frac{F^a}{Q^a}\Big)\Big|_{\bar D^+} \in
\Cinf(\bar D^+)\,,
\end{align}
where $\bT^+, \bT^-$ denote the extensions of $\bT$ to the boundary from $D^+$
and
$D^-$, respectively  (see the text above Theorem \ref{thm-app-uni}).
A normal to $N$, pointing from $D^-$ to $D^+$ is $n\equiv e_1$, the first standard
unit vector. Hence the projection of $\bT^+$ onto $n$, denoted by $\bT^+_n$, is
just its first component and the same holds true for  $\bT^-_n$. To apply
Corollary \ref{cor-app-uni} we have to show that ${\bT^+_n(\bW)>0}$ and
${\bT^-_n(\bW)>0}$ (for $\bW\in N$).

Since \eqref{eq-de} is obtained from rewriting the geodesic equation in first
order form we have
$E^1=0$ and $Q^1=1$, hence we obtain $\bT^+_n(\bW)=F^1(\bW)=\dot
U=\bT^-_n(\bW)$.
Now let $\bW(\tau)$ be a geodesic with initial value
$\bW(0)=(U_0,\dot U_0, V_0,\dot V_0,X_0,\dot X_0, Y_0, \dot Y_0)$ with $U_0
< 0$, which has $\dot{U}(\tau)\ge0$
before it reaches $N$.
(This is the only relevant case since if $U_0 < 0$ the geodesic can reach $N$
at all only if $\dot U\ge0$ just before it reaches $N$ and the cases
for $U_0 > 0$ are completely analogous.)
Assume that $\dot{U}=0$ at $N$, then reversing this geodesic, one obtains a
geodesic, which starts in $N$,
is tangential to $N$ and leaves it --- a contradiction to $N$ being totally geodesic.
Consequently ${\dot U>0}$ at $N$, hence ${\bT^+_n(\bW)>0}$, ${\bT^-_n(\bW)>0}$,
and we
have unique solutions in the sense off
Filippov by Corollary~\ref{cor-app-uni}.
\end{pr}

\subsection{The geodesic equation in nonexpanding impulsive waves}
\label{sec:3.4}

Our next goal is thus to explicitly write the geodesic equation for nonexpanding impulsive waves
in the continuous form and to see that we can apply Theorem \ref{mainthm}. The continuous line element
of nonexpanding impulsive waves is given by equation \eqref{conti}. Using the relation
\begin{equation}
Z=\frac{1}{\sqrt{2}}\big(X+iY\big) \,, \qquad\qquad \bar{Z}=\frac{1}{\sqrt{2}}\big(X-iY\big) \,, \label{RealContCoord}
\end{equation}
we obtain its real form
\begin{equation}
\dd{s}^2 = \omega^{-2}(U,V,X^k)\,\big[\,g_{ij}\,(U,X^k)\,\dd X^i \dd X^j
-2\,\dd U \dd V\big] \,,\label{eq-real-metric}
\end{equation}
${i,j=2,3}$ with $X^2\equiv X$, $X^3\equiv Y$,
where
\begin{eqnarray}
&& g_{ij} = \delta_{ij}+2\Up\, H_{,ij}+\Up^2\,\delta^{kl}H_{,ik}H_{,jl}\,, \label{Functiongij}\\
&& \omega = 1+\frac{\Lambda}{12}\big(\delta_{ij}X^i X^j -2UV-2\Up G\big)\,,\label{Functionomegaij}\\
&& G   =  H-X^iH_{,i} \,. \label{FunctionG}
\end{eqnarray}
Recall that by Rademacher's theorem, (locally) Lipschitz continuous functions
 are differentiable almost everywhere with derivative belonging (locally) to $L^\infty$.
Taking derivatives of the metric coefficients (e.g.\ $\Up$, $\Up^2$) will always be understood in this
sense. The Christoffel symbols for the metric \eqref{eq-real-metric} are
\begin{eqnarray}
&& \Gamma^U_{VV} = 0 \,, \qquad\qquad\quad\,\ \Gamma^U_{VU} = 0 \,, \qquad\qquad\qquad\,\ \Gamma^U_{Vi} = 0 \,, \nonumber \\
&& \Gamma^U_{UU} = -2\frac{\omega_{,U}}{\omega} \,, \qquad\quad \Gamma^U_{Ui} = -\frac{\omega_{,i}}{\omega} \,,
\quad\qquad\qquad\, \Gamma^U_{ij} = -g_{ij}\frac{\omega_{,V}}{\omega} \,, \label{NEx Christoffel u} \\
&& \ \nonumber \\
&& \Gamma^V_{VV} = -2\frac{\omega_{,V}}{\omega} \,, \qquad\quad\! \Gamma^V_{VU} = 0 \,, \qquad\quad\, \Gamma^V_{Vi} =
-\frac{\omega_{,i}}{\omega} \,, \nonumber \\
&& \Gamma^V_{UU} = 0 \,, \qquad\qquad\quad\,\ \Gamma^V_{Uj} = 0 \,, \qquad\quad\,\ \Gamma^V_{jk} =
-g_{jk}\frac{\omega_{,U}}{\omega}+\frac{1}{2}g_{jk,U} \,, \label{NEx Christoffel V} \\
&&\ \nonumber \\
&& \Gamma^i_{VV} = 0 \,, \qquad\qquad\quad\;\ \Gamma^i_{VU} = -g^{ij}\frac{\omega_{,j}}{\omega} \,, \qquad\quad\;\,
\Gamma^i_{Vk} = -\delta^i_k\frac{\omega_{,V}}{\omega} \,, \nonumber \\
&& \Gamma^i_{UU} = 0 \,, \qquad\qquad\quad\;\ \Gamma^i_{Uk} = \frac{1}{2}g^{ij}g_{jk,U}-\delta^i_k\frac{\omega_{,U}}{\omega}
\,, \nonumber \\
&& \Gamma^i_{kl} =
\,^{s}\Gamma^i_{kl}-\frac{1}{\omega}\left(\delta^i_k\omega_{,l}+\delta^i_l\omega_{,k}-g^{ij}g_{kl}\omega_{,j}\right) \,,
\label{NEx Christoffel i}
\end{eqnarray}
where $\,^{s}\Gamma^i_{kl}$ denotes the Christoffel symbols of the ``spatial'' metric $g_{ij}$.
The equations of geodesics in this case thus take the explicit form
\begin{eqnarray}\label{eq-geo-gen}
&& \ddot{U} -2\frac{\omega_{,U}}{\omega}\dot{U}^2 -g_{ij}\frac{\omega_{,V}}{\omega}\dot{X}^i\dot{X}^j
-2\frac{\omega_{,i}}{\omega}\dot{U}\dot{X}^i = 0 \,, \nonumber \\
&& \ddot{V} -2\frac{\omega_{,V}}{\omega}\dot{V}^2
-\Big(g_{jk}\frac{\omega_{,U}}{\omega}-\frac{1}{2}g_{jk,U}\Big)\dot{X}^j\dot{X}^k
-2\frac{\omega_{,i}}{\omega}\dot{V}\dot{X}^i = 0 \,,  \\
&& \ddot{X}^i
+\Big[\,^{s}\Gamma^i_{kl}-\frac{1}{\omega}\big(\delta^i_k\omega_{,l}+\delta^i_l\omega_{,k}-g^{ij}g_{kl}\omega_{,j}\big)\Big]
\dot{X}^k\dot{X}^l -2g^{ij}\frac{\omega_{,j}}{\omega}\dot{V}\dot{U} \nonumber \\
&& \hspace{5.0mm} -2\delta^i_k\frac{\omega_{,V}}{\omega}\dot{V}\dot{X}^k
+\Big(g^{ij}g_{jk,U}-2\delta^i_k\frac{\omega_{,U}}{\omega}\Big)\dot{U}\dot{X}^k = 0 \,.\nonumber
\end{eqnarray}

To explicitly see that \eqref{eq-geo-gen} is of the form \eqref{eq-de} we treat the cases $\Lambda=0$ and
$\Lambda\not=0$ separately.

\subsection{Geodesics in impulsive pp-waves}\label{mainlz}

In the {\it pp}-wave case, i.e., ${\Lambda=0}$ and hence ${\omega=1}$,
the geodesic equations \eqref{eq-geo-gen} simplify to
\begin{eqnarray}\label{eq-pp-geo}
\ddot{U} = 0 \,, \qquad\quad \ddot{V} +\frac{1}{2}g_{jk,U}\dot{X}^j\dot{X}^k = 0 \,, \qquad\quad  \ddot{X}^i
+\,^{s}\Gamma^i_{kl}\dot{X}^k\dot{X}^l +g^{ij}g_{jk,U}\dot{U}\dot{X}^k = 0 \,.
\end{eqnarray}

Recall that the main (technical) obstacle in case of general $\Lambda$ was that $U$ could not be used as an affine
parameter along the geodesics. However, in the case ${\Lambda=0}$ we have ${\ddot U=0}$ and hence $U$ \emph{can} be
used to parametrize the geodesics, significantly simplifying our task. Indeed, 
using $U$ as an affine parameter
and thus setting $\dot U=1$ the geodesic equations take the explicit form
\begin{align}\label{eq:geo-expl}
\ddot{V}=&
-\big[(U_+)_{,U}\,H_{,ij}+\textstyle{\frac{1}{2}}(U_+^2)_{,U}\,\delta^{mn}H_{,im}H_{,jn}\big]\dot{X}^i\dot{X}^j\,, \nonumber \\
\ddot{X}^i=&
-g^{ij}\big[U_+\,H_{,jkl}+U_+^2\,\delta^{mn}H_{,jm}H_{,kln}\big]\dot{X}^k\dot{X}^l \\
&-2g^{ij}\big[(U_+)_{,U}\,H_{,jk}+\textstyle{\frac{1}{2}}(U_+^2)_{,U}\,\delta^{mn}H_{,jm}H_{,kn}\big]\dot{X}^k \,,\nonumber
\end{align}
where the inverse spatial metric is given by
\begin{equation}
g^{ij}=D^{-1}g_{pq}(\delta^{ij}\delta^{pq}-\delta^{ip}\delta^{jq}\big)\,, \qquad
D\equiv\det g_{ij} =g_{22}g_{33}-(g_{23})^2\,.
\end{equation}
Now we extract $\Theta(U)$ from \eqref{eq:geo-expl} to put it into the form \eqref{eq-de}.
To this end we write the kinks as ${U_+=\Theta(U)U}$ and ${U_+^2=\Theta(U)U^2}$,
respectively, and (recalling that by \eqref{notation Xa} ${W^5=X=X^2}$,
${W^7=Y=X^3}$, ${W^6=\hat X=\dot X^2}$, and ${W^8=\hat Y=\dot X^3}$) we obtain

\begin{align}
 E^1 =&\  0,\ F^1 = \hat{U},\ Q^1 = 1\,,\nonumber
 \\
 E^2 =&\  0,\  F^2 = 0\,,\nonumber
 \\
 E^3 =&\  0,\ F^3 = \hat{V},\  Q^3 = 1\,,\nonumber
 \\
 E^4 =&\  -(H_{,ij} + U \delta^{kl} H_{,ik} H_{,jl})\hat{X}^i \hat{X}^j,\
 P^4 = 1,\  F^4 = 0\,,\nonumber
 \\
 E^a =&\  0,\ F^a = \hat{X}^i,\  Q^a = 1\,,\quad \mbox{for $a=5$
with $i=2$, and $a=7$ with $i=3$}\,,
 \\
 E^a =& -(\delta^{ij}\delta^{pq}-\delta^{ip}\delta^{jq})(\delta_{pq}
       + 2 U H_{,pq} + U^2\delta^{rs}H_{,pr}H_{,qs})\nonumber\\
&\times\big[(U H_{,jkl} + U^2\delta^{mn}H_{,jm}H_{,kln})\hat{X}^k\hat{X}^l + 2
(H_{,jk}+ U
\delta^{mn}H_{,jm}H_{,kn})\hat{X}^k\big]\nonumber\\
&\hspace*{38.5mm} \mbox{for $a=6$ with $i=2$, and $a=8$ with $i=3$}\,,\nonumber
 \\
 &F^6=0=F^8\,,\nonumber\\
 &P^6 =\  P^8 = D =
\big[1+(H_{,22}+H_{,33})\,\Up+(H_{,22}H_{,33}-H_{,23}^2)\,\Up^2\big]^{2}\,.\nonumber
\end{align}
Of course, $P^a$ is irrelevant whenever ${E^a=0}$, and similarly $Q^a$
whenever ${F^a=0}$.

Since in addition all Christoffel symbols of the form $\Gamma^U_{\mu\nu}$
vanish, the hyperplane ${\{U=0\}}$ is totally geodesic and we may apply
Theorem \ref{mainthm}. It thus follows that, provided
$H$ is smooth, the geodesic equation for data with $U\ne0$ in impulsive {\it
pp}-waves is uniquely solvable in the sense of Filippov, and the geodesics are
$\Con^1$-curves. Those which hit the wave surface also cross it. Since
the background spacetime off the shock surface, i.e., Minkowski space is clearly
complete we also have a completeness result. Note, however, that we do not
obtain any additional information on the geodesics which lie within the null surface
$N$.
\begin{thm}\label{ppthm}
 Consider the impulsive {\it pp}-wave spacetime \eqref{conti} with
 ${\Lambda=0}$ and smooth $H$. Given initial data off the wave surface ${\{U=0\}}$
the geodesic equation possesses global unique \hbox{$\Con^1$-geodesics} (in the sense
of Filippov). In particular, these geodesics are complete.
\end{thm}

This theorem applies e.g.\ to plane waves, where $H$ is quadratic in $X$, $Y$, 
as well as to \emph{pp}-waves with $H$ being a higher polynomial \cite{[C2]}.
However, in many physically interesting models $H$ will be non-smooth, possessing
poles at the axis ${\{\rho=0\}}$, cf.\ section \ref{sec:boost}.
In such a case Theorem \ref{mainthm} can still be applied, but some
care is needed. Indeed, if a geodesic starts in the (background) region where
${U<0}$ and hits the wave surface at ${U=0}$ with some ${\rho(U=0)>0}$ we may work on
an open subset of the spacetime with a small neighborhood of ${\rho=0}$, ${U\geq 0}$
removed. There the metric is locally Lipschitz and Theorem \ref{mainthm} still
applies to guarantee that the geodesic continues into the region ${U>0}$ and
exists as a unique, $\Con^1$- solution in the sense of Filippov at least for
small positive $U$. Of course, later on it might run into the singularity at
${U>0}$, ${\rho=0}$ which, however, is just a coordinate singularity introduced in
the Minkowski background via the transformation \eqref{ro:trsf}. Hence the
geodesic will be complete. Also we can argue analogously for geodesics starting
with positive $U$ and running towards the wave surface. The only geodesics which
do \emph{not} allow for such an application of Theorem \ref{mainthm} are those
which \emph{directly head} at the curvature singularity at ${U=0}$, ${\rho=0}$, which is in
complete agreement with physical expectations. Summing up we have the following
result:

\begin{thm}\label{ppthm1}
 Consider the impulsive {\it pp}-wave spacetime \eqref{conti} with
 ${\Lambda=0}$ and $H$ smooth off ${\rho=0}$. Then given initial
 data off the wave surface ${\{U=0\}}$ the geodesic equation possesses locally
 defined unique $\Con^1$-solutions (in the sense of Filippov). Moreover, all geodesics
starting at ${U\not=0}$ and not directly heading towards ${U=0}$, ${\rho=0}$ are
complete.
\end{thm}

 As already said in section \ref{sec:3.2} the fact that $U$ can serve as an
affine parameter along the geodesics
 in impulsive {\it pp}-waves makes it possible to employ the (simpler) solution
concept of
 Carath\'{e}odory \cite[Chapter 1]{F:88}, see \cite{LSS:13}. Since in general
Filippov and Carath\'{e}odory
 solutions do not agree we face the question of compatibility of Theorem
\ref{ppthm} with
 the result in \cite{LSS:13}. However, it is easily seen (from the fact that for
 $U\neq 0$ the Filippov set-valued map satisfies $\mathcal{F}[f](U)=\{f(U)\}$,
where $f$ is the right hand-side of
 the ODE) that for the equations \eqref{eq-pp-geo} any Carath\'{e}odory solution
is a Filippov solution
 and vice versa. Hence the geodesics agree in both approaches, a fact that is
also clearly visible from the explicit junction conditions
\eqref{RefForm} derived below in section \ref{sec:matching}.

\subsection{Geodesics in the general case with \texorpdfstring{${\Lambda\neq0}$}{Lambda not zero}}
\label{mainlnz}

With any value of the cosmological constant $\Lambda$, the function ${\omega}$
takes the general form \eqref{Functionomegaij}. The corresponding
equations of geodesics \eqref{eq-geo-gen}, representing motion caused by
impulsive gravitational waves propagating in (anti-)de~Sitter universe, are thus
given by a considerably more complex system than \eqref{eq-pp-geo} valid in flat
Minkowski background. Nevertheless, it is again possible to put
\eqref{eq-geo-gen} into the form \eqref{eq-de}. A straightforward but somewhat
lengthy calculation reveals that
\begin{align}
E^1 &= 0,\ {F}^1 = \hat{U},\ Q^1 = 1\,,
\nonumber\\
E^2 &= -\frac{\Lambda}{3}\left[G\,\hat{U}^2+\big( H_{,ij} +\frac{1}{2}
U\delta^{kl}H_{,ik}H_{,jl}\big)U^2\hat{X}^i\hat{X}^j
-U\hat{U}H_{,ki}\,X^k\hat{X}^i\right],\
P^2 = \omega\,, \nonumber\\
&{F}^2 = -\frac{\Lambda}{3}\left[V\hat{U}^2 +\frac{1}{2} U
\delta_{ij}\hat{X}^i \hat{X}^j-\hat{U} \delta_{ik}\hat{X}^i X^k\right],\
Q^2 = \omega\,, \nonumber\\
E^3 &= 0,\ {F}^3 = \hat{V},\ Q^3 = 1\,,
\nonumber\\
E^4 &= -\bigg[\big(H_{,ij} + U \delta^{kl} H_{,ik} H_{,jl}\big)
\Big(1+\frac{\Lambda}{12}\delta_{mn}X^mX^n\Big)+\frac{\Lambda}{6}\Big(G\delta_{
ij}+U(V+G)H_{,ij}\Big)\bigg]\hat{X}^i\hat{X}^j\nonumber\\
&\qquad
+\frac{\Lambda}{3}U\hat{V}H_{,ki}\,X^k\hat{X}^i,\
P^4 = \omega\,, \nonumber\\
&{F}^4 = -\frac{\Lambda}{3}\left[U\hat{V}^2 +\frac{1}{2} V
\delta_{ij}\hat{X}^i \hat{X}^j-\hat{V} \delta_{ik}\hat{X}^i X^k\right],\
Q^4 = \omega\,, \nonumber\\
E^a &= 0,\ {F}^a = \hat{X}^i,\ Q^a = 1\,,\quad \mbox{for $a=5$
with $i=2$, and $a=7$ with $i=3$}\,, \label{Ea} \\
E^a &= -(\delta^{ij}\delta^{pq}-\delta^{ip}\delta^{jq})\, g_{pq}^+ \nonumber\\
&\ \times\bigg\{\Big[\omega_+(U H_{,jkl} + U^2\delta^{mn}H_{,jm}H_{,kln})
+\frac{\Lambda}{6}\big((2UH_{,kl}+ U^2 \delta^{mn}H_{,km}H_{,ln})\delta_{js}X^s
-G_{,j}g_{kl}^+U\big)
\Big]\hat{X}^k\hat{X}^l \nonumber\\
&\qquad +2\omega_+(H_{,jk}+ U \delta^{mn}H_{,jm}H_{,kn})\hat{U}\hat{X}^k
+\frac{\Lambda}{3}G_{,j}U\hat{U}\hat{V}\bigg\}\nonumber\\
&\ -\frac{\Lambda}{6}(2UH_{,pq}+ U^2 \delta^{mn}H_{,pm}H_{,qn})
(\delta^{pq}X^i-\delta^{ip}X^q)
(\delta_{kl}\hat{X}^k\hat{X}^l-2\hat{U}\hat{V})\nonumber\\
&\ +\frac{\Lambda}{3}\bigg\{U H_{,kl}X^k\hat{X}^l-G\hat{U}+{\mathcal D}\Big[
\delta_{jk}X^j\hat{X}^k-(U\hat{V}+\hat{U}V)+U H_{,kl}X^k\hat{X}^l
-G\hat{U}\Big]\bigg\}\hat{X}^i \nonumber\\
&\hspace*{4.2cm} \mbox{for $a=6$ with $i=2$, and $a=8$ with
$i=3$}\,,\nonumber
\end{align}
\begin{align}
&{F}^a =\frac{\Lambda}{3}\left[X^i\Big(\hat{U}\hat{V}-\frac{1}{2}\delta_{kl}
\hat {X }^k\hat{X}^l\Big)
+\hat{X}^i\Big(\delta_{jk}X^j\hat{X}^k-(U\hat{V}+\hat{U}V)\Big)\right]
\nonumber\\
&\hspace*{4.2cm}\mbox{for $a=6$ with $i=2$, and $a=8$ with
$i=3$}\,,
\nonumber\\
&P^6 = P^8 =Q^6=Q^8= D\,\omega\,,\nonumber
\end{align}
where
\begin{align}
 g_{ij}^+ &= \delta_{ij} + 2 U H_{,ij} + U^2\delta^{kl}H_{,ik}H_{,jl}\,,
\nonumber\\
 \omega_+ &= 1+\frac{\Lambda}{12}\big(\delta_{ij}X^i X^j -2UV-2UG\big), \\
 {\mathcal D} &= \ \big[(H_{,22}+H_{,33})+U(H_{,22}H_{,33}-H_{,23}^2)\big]
 \,
 \big[2U+U^2(H_{,22}+H_{,33})+U^3(H_{,22}H_{,33}-H_{,23}^2)\big]\,,
\nonumber
\end{align}
are smooth functions, polynomial in $U$. Of course, for ${\Lambda=0}$ these
expressions reduce to those presented in the previous section.

Also, the null hypersurface given by $U=0$ is again 
totally geodesic: $\omega_{,V}$ is proportional to $U$, and so the geodesic equation
\eqref{eq-geo-gen} with initial data $U=0$ and $\dot U=0$ allows for the solution $U\equiv0$.
By uniqueness this implies that $\{U=0\}$ is totally geodesic in the background (anti-)de Sitter spacetime.
Suppose now that in the impulsive wave spacetime such a geodesic leaves
the hypersurface for $\tau>\tau_0$, it also would be a geodesic in the background for $\tau>\tau_0$.
However, by continuity of the tangent vector this implies $\dot U(\tau_0)=0$ for a geodesic in the background, 
contradicting the above.

So for smooth $H$ we obtain a result directly generalizing Theorem \ref{ppthm}:

\begin{thm}\label{genthm}
 Consider the nonexpanding impulsive wave spacetime \eqref{conti} with
 arbitrary ${\Lambda}$ and smooth $H$. Given initial data off the wave
surface ${\{U=0\}}$ the geodesic equation possesses global unique
\hbox{$\Con^1$-geodesics} (in the sense of
 Filippov). In particular, these geodesics are complete.
\end{thm}

Also the physically more relevant models with singular $H$ can be dealt with similar to
the case of vanishing $\Lambda$. Observe from \eqref{E4.5} that in the present case the singularities 
are localized on the wave front at ${z=\pm 1}$. Hence we also obtain a generalization of Theorem \ref{ppthm1}:

\begin{thm}\label{genthm1}
 Consider the nonexpanding impulsive wave spacetime \eqref{conti} with
 arbitrary ${\Lambda}$ and~$H$ smooth off ${z=\pm 1}$. Then given
 initial data off the wave surface ${\{U=0\}}$ the geodesic equation possesses
 locally defined unique \hbox{$\Con^1$-solutions} (in the sense of Filippov). Moreover, all
 geodesics starting with ${U\not=0}$ and not directly heading towards ${U=0}$, ${z=\pm 1}$ are complete.
\end{thm}

In view of the discussion in section \ref{subsec:emb} the two singularities occur at ${Z_2=0=Z_3}$ and
${Z_4=\pm a}$. These are the north and south poles of a spherical impulsive wave in de Sitter space and
the vertices of hyperboloidal waves in anti-de Sitter space.

\section{\texorpdfstring{The
$\Con^1$-matching}{C1-matching}}\label{sec:matching}

In this section we apply the results of sections~\ref{mainlz} and \ref{mainlnz} to
explicitly derive the form of the geodesics in nonexpanding impulsive
gravitational waves by appropriatetly matching the geodesics of the background. We, however, start with a general remark.

\begin{rem}[The philosophy of the matching]\leavevmode
\begin{enumerate}
 \item Observe that the matching is only justified \emph{after} we have gained sufficient knowledge on the geodesics of 
  the entire spacetime: the geodesics heading
  towards the wave surface \emph{cross} it, are \emph{unique} and of \emph{$\Con^1$-regularity}.
 \item However, one may consider the following more general situation where such a procedure is possible:
  Assume we have a $\Con^{0,1}$-metric and the spacetime is separated by a hypersurface $N$ into 
  two parts $D^+$ and $D^-$ such that $g|_{D^\pm}\in\Con^\infty$. Then, in
  particular, one has (unique) classical (smooth) geodesics on both sides. Now provided the right hand side of the 
  geodesic equation (written as a first order system, cf.\ \eqref{eq-de}) satisfies $\bT^\pm_n>0$, these geodesics combine to unique ($\Con^1$-)solutions in the sense of Fillipov. 
  Hence the global geodesics can be computed simply by matching the background geodesics in a $\Con^1$-manner without the 
  need to go into the details of Filippov's theory.
\end{enumerate}
\end{rem}

To explicitly carry out the $\Con^1$-matching procedure in our case we start with the 
unique globally defined $\Con^1$-geodesics in the continuous metric \eqref{eq-real-metric}.
Transforming them into coordinate 
systems~\eqref{conf*} well adapted to the background spacetimes separately on either side of the wave surface we derive
explicit matching conditions for the position and the velocity of the geodesics
of the background across ${\{U=0\}}$.

To begin with we write the background spacetimes of constant curvature (\ref{conf*}) in the real coordinates
\begin{equation}
x=\frac{1}{\sqrt{2}}(\eta+\bar{\eta}) \,, \qquad y=\frac{1}{\sqrt{2}\,i}(\eta-\bar{\eta}) \,. \label{CoordReal}
\end{equation}
Using the real spatial variables (\ref{RealContCoord}) and (\ref{CoordReal}), the transformation (\ref{trans}), which relates the continuous line element (\ref{conti}) to the constant curvature background spacetimes (\ref{conf*}), can be expressed~as
\begin{eqnarray}
&& \U = U \,, \nonumber \\
&& \V = V+\Theta H+\textstyle{\frac{1}{2}}\Up\big((H_{,X})^2+(H_{,Y})^2\big) \,, \nonumber \\
&& x = X+\Up\,H_{,X} \,, \nonumber \\
&& y = Y+\Up\,H_{,Y} \,, \label{DiscontTransReal}
\end{eqnarray}
where ${H=H(X,Y)}$.  Now we consider geodesics
\begin{eqnarray}
U=U(\tau) \,, \qquad V=V(\tau) \,, \qquad X=X(\tau) \,, \qquad Y=Y(\tau) \,, \label{GeodesicsReal}
\end{eqnarray}
in the continuous metric \eqref{eq-real-metric}.
By the results of section \ref{mainlnz} they are unique globally defined $\Con^1$-curves. In particular,
positions and velocities at the instant of interaction with the impulse are \emph{equal on both sides}.
Hence by employing the transformation (\ref{DiscontTransReal}) and its derivative separately in
the region ${U>0}$ and ${U<0}$ we can express the refraction formulae for the
geodesics crossing the impulsive hypersurface ${U=0}$ as
\begin{eqnarray}
&& \U_{\mathrm{i}}^- = 0 = \U_{\mathrm{i}}^+ \,, \qquad\quad
\dot{\U}^-_{\mathrm{i}}=\dot{\U}^+_{\mathrm{i}} \,, \nonumber \\
&& \V_{\mathrm{i}}^- = \V_{\mathrm{i}}^+-H_{\mathrm{i}} \,, \qquad\;\
\dot{\V}^-_{\mathrm{i}}=\dot{\V}^+_{\mathrm{i}}-H_{\mathrm{i},X}\,\dot{x}_{
\mathrm{i}}^+-H_{\mathrm{i},Y}\,\dot{y}_{\mathrm{i}}^++\textstyle{\frac{1}{2}}
\big((H_{\mathrm{i},X})^2+(H_{\mathrm{i},Y})^2\big)\,\dot{\U}_{\mathrm{i}}^+ \,,
\label{RefForm} \\
&& x_{\mathrm{i}}^- = x_{\mathrm{i}}^+ \,, \qquad\qquad\quad\
\dot{x}^-_{\mathrm{i}}=\dot{x}^+_{\mathrm{i}}-H_{\mathrm{i},X}\,\dot{\U}_{
\mathrm{i}}^+ \,, \nonumber \\
&& y_{\mathrm{i}}^- = y_{\mathrm{i}}^+ \,, \qquad\qquad\quad\,\
\dot{y}^-_{\mathrm{i}}=\dot{y}^+_{\mathrm{i}}-H_{\mathrm{i},Y}\,\dot{\U}_{
\mathrm{i}}^+ \,. \nonumber
\end{eqnarray}
Here the subscript $_\mathrm{i}$ denotes the values of the respective quantities at the instant when the geodesics 
interact with the impulse at ${U=0}$ (note that ${H_{\mathrm{i},X}=(H_{,X})_\mathrm{i}}$), while the superscripts $+$ and $-$ denote the values of the positions and
velocities of the geodesics as they approach the impulse from the region
${\{U<0\}}$ resp.\ ${\{U>0\}}$. Interestingly, these relations do not explicitly depend on the cosmological
constant $\Lambda$ since the conformally flat coordinates are used. Moreover they clearly reduce to the
conditions derived in \cite{LSS:13} in the case ${\Lambda=0}$.

However, to better understand the influence of the (anti-)de Sitter background, it is
convenient to employ the 5-dimensional formalism (see
section~\ref{subsec:emb} and~\cite{PO:01}). Specifically, we can work with the metric \eqref{general} which is related to (\ref{confppimp}) 
by transformation \eqref{Zcoord}.

Defining the evaluation of the conformal factor on either side by 
$\Omega_{\mathrm{i}}^\pm=1+{\textstyle\frac{1}{12}}\Lambda\big((x^{\pm}_{\mathrm{i}})^2+(y^{\pm}_{\mathrm
{i}})^2\big)=2a/(Z_{4\mathrm{i}}^{\pm}+a)$ we find, using the fact 
${x_{\mathrm{i}}^- = x_{\mathrm{i}}^+}$ and ${y_{\mathrm{i}}^- = y_{\mathrm{i}}^+}$, that 
${\Omega_{\mathrm{i}}^{-}=\Omega_{\mathrm{i}}^{+}}$  and we may just denote it as 
$\Omega_{\mathrm{i}}$. So we obtain from \eqref{RefForm}
\begin{equation}
\tilde{U}_{\mathrm{i}}^- = 0 = \tilde{U}_{\mathrm{i}}^+ \,, \qquad\   \tilde{V}_{\mathrm{i}}^- = \tilde{V}_{\mathrm{i}}^+-\frac{H_{\mathrm{i}}}{\Omega_{\mathrm{i}}} \,,
\qquad Z_{2\mathrm{i}}^- = Z_{2\mathrm{i}}^+\,,\qquad Z_{3\mathrm{i}}^- = Z_{3\mathrm{i}}^+\,,  \qquad
Z_{4\mathrm{i}}^- = Z_{4\mathrm{i}}^+\,, \label{5DRefPosNull}
\end{equation}
which are in fact the Penrose junction conditions \eqref{juncnon} in $5$ dimensions. 

Moreover, ${\dot{\Omega}^{-}_{\mathrm{i}}={\dot{\Omega}^{+}_{\mathrm{i}}}+\frac{1}{2\epsilon a^2}\,G_{\mathrm
{i}}\,\Omega_{\mathrm{i}}\,\dot{\tilde{U}}^{+}_{\mathrm{i}}}$ where for 
${G_{\mathrm{i}}=G^{\pm}_{\mathrm{i}}=H_{\mathrm{i}}-H_{\mathrm{i},X}\,\Omega_{
\mathrm{i}}\,Z^{\pm}_{2\mathrm{i}}-H_{\mathrm{i},Y}\,\Omega_{\mathrm{i}}\,Z^{\pm
}_{3\mathrm{i}}}$ (see \eqref{FunctionG} with \eqref{DiscontTransReal} and \eqref{Zcoord}) we have adopted a convention analogous to that for $\Omega_{\mathrm{i}}$.
In this way we obtain for the velocities 
\begin{eqnarray}
&& \dot{\tilde{U}}^{-}_{\mathrm{i}} = \dot{\tilde{U}}^{+}_{\mathrm{i}} \,, \nonumber\\
&& \dot{\tilde{V}}^{-}_{\mathrm{i}} = \dot{\tilde{V}}^{+}_{\mathrm{i}}+2p\,\dot{\tilde{U}}^{+}_{\mathrm{i}} -H_{\mathrm{i},X}\dot{Z}^{+}_{2\mathrm{i}}-H_{\mathrm{i},Y}\dot{Z}^{+}_{3\mathrm{i}}-\frac{G_{\mathrm{i}}}{2a}\,\dot{Z}^{+}_{4\mathrm{i}} \,,  \nonumber\\
&& \dot{Z}^{-}_{2\mathrm{i}} = \dot{Z}^{+}_{2\mathrm{i}} -\dot{\tilde{U}}^{+}_{\mathrm{i}}\Big(H_{\mathrm{i},X}+\frac{G_{\mathrm{i}}}{2\epsilon a^2}\,Z_{2\mathrm{i}}^{+}\Big) \,, \label{5DRefPosNull1}\\
&& \dot{Z}^{-}_{3\mathrm{i}} = \dot{Z}^{+}_{3\mathrm{i}} -\dot{\tilde{U}}^{+}_{\mathrm{i}}\Big(H_{\mathrm{i},Y}+\frac{G_{\mathrm{i}}}{2\epsilon a^2}\,Z_{3\mathrm{i}}^{+}\Big) \,, \nonumber \\
&& \dot{Z}^{-}_{4\mathrm{i}} = \dot{Z}^{+}_{4\mathrm{i}}-\dot{\tilde{U}}^{+}_{\mathrm{i}}\frac{G_{\mathrm{i}}}{\epsilon a\Omega_{\mathrm{i}}}  \nonumber\,,
\end{eqnarray}
where 
\begin{equation}
p \equiv \frac{1}{4}\Big[(H_{\mathrm{i},X})^2+(H_{\mathrm{i},Y})^2-\frac{G_{\mathrm{i}}}{\epsilon a^2}\,\Big(\tilde{V}^{+}_{\mathrm{i}}-\frac{H_{\mathrm{i}}}{\Omega_{\mathrm{i}}}\Big)\Big] \,.
\end{equation}

The above expressions can also be rewritten in the $5$-dimensional Minkowski coordinates as
\begin{equation}
Z_{0\mathrm{i}}^- = Z_{0\mathrm{i}}^+-\frac{H_{\mathrm{i}}}{\sqrt2\,\Omega_{\mathrm{i}}} \,, \qquad\   Z_{1\mathrm{i}}^- = Z_{1\mathrm{i}}^++\frac{H_{\mathrm{i}}}{\sqrt2\,\Omega_{\mathrm{i}}} \,,
\qquad Z_{2\mathrm{i}}^- = Z_{2\mathrm{i}}^+\,,\qquad Z_{3\mathrm{i}}^- = Z_{3\mathrm{i}}^+\,,  \qquad
Z_{4\mathrm{i}}^- = Z_{4\mathrm{i}}^+\,, \label{5DRefPos}
\end{equation}
for the positions and
\begin{eqnarray}
&& \dot{Z}^{-}_{0\mathrm{i}} = (1+p)\,\dot{Z}^{+}_{0\mathrm{i}}+p\,\dot{Z}^{+}_{1\mathrm{i}} -\frac{1}{\sqrt{2}}\Big(H_{\mathrm{i},X}\dot{Z}^{+}_{2\mathrm{i}}+H_{\mathrm{i},Y}\dot{Z}^{+}_{3\mathrm{i}}+\frac{G_{\mathrm{i}}}{2a}\,\dot{Z}^{+}_{4\mathrm{i}}\Big) \,,  \nonumber\\
&& \dot{Z}^{-}_{1\mathrm{i}} = -p\,\dot{Z}^{+}_{0\mathrm{i}}+(1-p)\,\dot{Z}^{+}_{1\mathrm{i}} +\frac{1}{\sqrt{2}}\Big(H_{\mathrm{i},X}\dot{Z}^{+}_{2\mathrm{i}}+H_{\mathrm{i},Y}\dot{Z}^{+}_{3\mathrm{i}}+\frac{G_{\mathrm{i}}}{2a}\,\dot{Z}^{+}_{4\mathrm{i}}\Big) \,,  \nonumber\\
&& \dot{Z}^{-}_{2\mathrm{i}} = \dot{Z}^{+}_{2\mathrm{i}} -\frac{1}{\sqrt{2}}\big(\dot{Z}^{+}_{0\mathrm{i}}+\dot{Z}^{+}_{1\mathrm{i}}\big)\Big(H_{\mathrm{i},X}+\frac{G_{\mathrm{i}}}{2\epsilon a^2}\,Z_{2\mathrm{i}}^{+}\Big) \,,\label{5DRefVel}\\
&& \dot{Z}^{-}_{3\mathrm{i}} = \dot{Z}^{+}_{3\mathrm{i}} -\frac{1}{\sqrt{2}}\big(\dot{Z}^{+}_{0\mathrm{i}}+\dot{Z}^{+}_{1\mathrm{i}}\big)\Big(H_{\mathrm{i},Y}+\frac{G_{\mathrm{i}}}{2\epsilon a^2}\,Z_{3\mathrm{i}}^{+}\Big) \,,  \nonumber\\
&& \dot{Z}^{-}_{4\mathrm{i}} = \dot{Z}^{+}_{4\mathrm{i}}-\frac{1}{\sqrt{2}}\big(\dot{Z}^{+}_{0\mathrm{i}}+\dot{Z}^{+}_{1\mathrm{i}}\big)\frac{G_{\mathrm{i}}}{\epsilon a\Omega_{\mathrm{i}}}  \nonumber\,,
\end{eqnarray}
for the velocities, where
\begin{equation}
p \equiv \frac{1}{4}\Big[(H_{\mathrm{i},X})^2+(H_{\mathrm{i},Y})^2+\frac{\sqrt{2}G_{\mathrm{i}}}{\epsilon a^2}\,\Big(Z_{1\mathrm{i}}^{+}+\frac{H_{\mathrm{i}}}{\sqrt{2}\,\Omega_{\mathrm{i}}}\Big)\Big] \,.
\end{equation}

The matching conditions derived above clearly demonstrate the following behaviour of the geodesics: 
\emph{As seen in the ``halves'' of the background (anti-)de Sitter spacetimes}
in front and behind the impulsive wave, they are \emph{refracted} in all directions but the one normal to the wave surface, 
see \eqref{RefForm}, \eqref{5DRefPosNull1}, \eqref{5DRefVel}. Additionally, they suffer a \emph{jump} in the ${\mathcal V}$-coordinate \eqref{RefForm} 
(respectively the ${\tilde V}$-coordinate \eqref{5DRefPosNull}, respectively the corresponding $Z_0$- and $Z_1$-components
\eqref{5DRefPos}).

\section*{Acknowledgment}
We thank Alexander Lecke for participating in our discussions in the early stage of this project.

JP and R\v{S} were supported by the Albert Einstein Center for Gravitation and Astrophysics, 
Czech Science Foundation 14-37086G, project
UNCE~204020/2012 and the grant 7AMB13AT003 of the Scientific and Technological
Co-operation Programme Austria--Czech Republic. CS and RS \hbox{acknowledge} the support of
its Austrian counterpart, OEAD's WTZ grant CZ15/2013 and of FWF grant P25326.

\begin{appendix}
\section*{Appendix A: Filippov solutions}\label{app:a}
\addcontentsline{toc}{section}{Appendix A: Filippov solutions}
\setcounter{section}{1}
\setcounter{thm}{0}
\def\thesection{\Alph{section}}

In this appendix we briefly recall the essentials of Filippov's solution
concept \cite{F:88} for ordinary differential equations with discontinuous
right hand side. Due to our exclusive interest in geodesic equations we only
discuss the autonomous case. For a general pedagogical introduction into the
topic see e.g.\ \cite{C:08}.

Consider a $d$-dimensional system of the form
\begin{equation}\label{eq-ode}
 \dot x(t) = f(x(t))\qquad (t\in I)\,,
\end{equation}
where $I$ is some interval in $\R$, $D\subseteq \R^d$ and $f\colon D\rightarrow \R^d$
is given. Peano's classical existence theorem needs the right hand side $f$ to
be continuous. Without this assumption, however, one has to change the
solution concept since then $x$ is not continuously differentiable in
general. The idea of Filippov's approach is to average the values of the
right hand side in a neighborhood of points of discontinuity. Formally one
associates to the right hand side $f$ a set-valued map $\mathcal{F}[f]: D
\rightarrow \mathcal{B}(D)$ (the collection of all
non-empty, closed and convex subsets of $D$) via
\begin{equation*}
 \mathcal{F}[f](x)\equiv\bigcap_{\delta>0}\bigcap_{\mu(S)=0}
\mathrm{co}(f(B_\delta(x)\backslash S))\,.
\end{equation*}
Here $\mathrm{co}(A)$ denotes the closed convex \emph{hull} of a set $A$, i.e., the smallest convex set
containing $A$. Moreover, $B_\delta(x)$ denotes the
closed Euclidean \emph{ball} around $x$ of radius $\delta$, and $\mu$ is the
Lebesgue measure on $\R^d$. Hence $\mathcal{F}[f](x)$ is given as the
intersection of convex hulls of images of
shrinking closed balls around $x$, while ignoring sets $S$ of measure zero. Observe
that the value of $\mathcal{F}[f]$ at $x$ is independent of the value of $f$ at
$x$, but at points of continuity we have $\mathcal{F}[f](x)=\{f(x)\}$.
Finally one replaces the differential equation \eqref{eq-ode} by the
\emph{differential inclusion}
\begin{equation}\label{eq-di}
 \dot x(t) \in \mathcal{F}[f](x(t))\,,
\end{equation}
in this way prescribing a \emph{range of values} for $\dot x$ rather than a single
value. Now we are in the position to define the notion of a Filippov solution.

\begin{defi}
 A Filippov solution of \eqref{eq-ode} on an interval $[a,b]\subseteq
I$ is an absolutely continuous curve $x:[a,b]\rightarrow D$, that satisfies
\eqref{eq-di} almost everywhere.
\end{defi}

Recall that a curve ${x:[a,b]\rightarrow \R^d}$ is said to be \emph{absolutely continuous} if for every ${\ep>0}$ there is a
${\delta>0}$ such that for all collections of non-overlapping intervals
${([a_i,b_i])_{i=1}^m}$ in ${[a,b]}$ with ${\sum_{i=1}^m(b_i-a_i)<\delta}$ we have
that ${\sum_{i=1}^m \|x(b_i)-x(a_i)\|< \ep}$. Moreover, recall that an absolutely
continuous curve is differentiable almost everywhere.

Of course, any classical, i.e., ${\mathcal C}^2$-solution is a Filippov solution but the latter exist under much more
general conditions. In fact, Filippov in \cite{F:88} has developed a complete theory of ordinary
differential equations based on this solution concept which has been found to be
widely applicable e.g.\ in non-smooth mechanics. Here we just state two simple
results suitable for our purpose.

\begin{thm}(Existence, \cite[Theorem 7.8]{F:88})
If ${f:D\rightarrow \R^d}$ is bounded, then for each $(t_0,x_0)\in I\times D$
there is a Filippov solution $x$ of \eqref{eq-ode} with $x(t_0)=x_0$.
\end{thm}

Just to give a simple example we consider an ODE with the Heaviside function as right hand side.
\begin{ex}\label{ex-heav}
 Let $\Theta\in L^\infty(\R)$ denote the Heaviside function, defined by
$\Theta(u)=0$ if $u<0$ and $\Theta(u)=1$ for $u>0$.
Recall that $\Theta$ as a class in $L^\infty(\R)$ has no value assigned at $0$
and the Filippov set-valued map does not depend on the value of the function at
a single point. So we easily obtain
\begin{equation*}
 \mathcal{F}[\Theta](u)=\begin{cases}
                         \{0\} 	& u<0\,,\\
                         [0,1] & u=0\,,\\
                         \{1\} 	& u>0\,,
                        \end{cases}
\end{equation*}
since $\Theta$ is continuous at $u\neq 0$ and $[0,1]$ is the smallest closed,
convex set containing $0$ and $1$.

We now consider the differential inclusion $\dot{u}(t)\in\mathcal{F}[\Theta](u(t))$
with initial condition $u(0)=u_0$. If $u_0<0$, then $u(t):= u_0$ $(t\in[0,\infty))$
is the unique Filippov solution with this initial condition. Similarly, if $u_0>0$,
then $u(t):= u_0 + t$ $(t\in[0,\infty))$ is the unique Filippov solution. If $u_0=0$,
however, the functions $u_1(t):= 0$, $u_2(t):=t$ $(t\in[0,\infty))$ are both
Filippov solutions starting at $0$.
\end{ex}

Uniqueness, in general, is more difficult to achieve. Recall that already
classical uniqueness theorems use a Lipschitz condition. While one-sided
Lipschitz conditions can be used to prove one-sided uniqueness of Filippov
solutions \cite[Sec.\ 10.1]{F:88}, they turn out to be ill-suited for
piecewise continuous right hand sides (cf.\ e.g.\ \cite[p.\ 53]{C:08}), which
are our main interest. Hence we will resort to results derived in \cite[Sec.\
10.2]{F:88} and assume that $D\subseteq \R^d$ is connected and separated by a
smooth surface $N$ into two domains $D^+$ and $D^-$. Let $f$ and $\pder{f}{x^i}$
($i=1,\ldots,d$) be continuous in $D^+$ and $D^-$
up to the boundary $N$. Denote by $f^+$ (respectively $f^-$) the extensions of
$f|_{D^+}$ (respectively $f|_{D^-}$) to the
boundary. Then set $h(x)\equiv f^+(x)-f^-(x)$ for $x\in N$ and let $f^+_n$, $f^-_n$,
$h_n$ be the projections of $f^+,f^-, h$ onto
the normal to $N$ directed from $D^-$ to $D^+$ at the points of $N$.

\begin{thm}(Sufficient conditions for uniqueness,
\cite[Lemma 10.2]{F:88})\label{thm-app-uni} If for $x_0\in N$ we have
$f^+_n(x_0)>0$, then in the domain $D^+$ there exists a unique Filippov solution
of \eqref{eq-ode} starting at~$x_0$. Analogous assertions hold for $D^-$ and
$f^-_n(x_0)<0$.
\end{thm}

Actually we will make use of the following result.
\begin{cor}(\cite[Corollary 10.1]{F:88}\label{cor-app-uni}
 On the region of the surface $N$ where ${f^+_n>0}$ and ${f^-_n>0}$ the solutions pass
from $D^-$ to $D^+$ and uniqueness is not violated.
\end{cor}

We conclude this appendix with an example and a remark relevant to our work.
\begin{ex}
 We consider the following variant of Example \ref{ex-heav}: Let
 \begin{equation*}
  f(u)=\begin{cases}
         1 & u<0\,,\\
         2 & u>0\,,
        \end{cases}\quad
        \text{ then }\quad
   \mathcal{F}[f](u)=\begin{cases}
                         \{1\} 	& u<0\,,\\
                         [1,2] & u=0\,,\\
                         \{2\} 	& u>0\,.
                        \end{cases}
 \end{equation*}
We consider $\dot{u}(t)\in\mathcal{F}[f](u(t))$ with $u(0)=u_0<0$, then
\begin{equation*}
 u(t)=\begin{cases}
    u_0 + t & t\leq -u_0\,,\\
    2(u_0 + t) & t\geq -u_0\,,
    \end{cases}
\end{equation*}
is the unique Filippov solution by Corollary \ref{cor-app-uni}. Indeed, we have $D^-=(-\infty,0)$, $D^+=(0,\infty)$, $N=\{0\}$
and $f_n^- = f^- = 1$, $f_n^+ = f^+ = 2>0$.
\end{ex}

For a general function $f$ it may be difficult to calculate the Filippov
associated
map $\mathcal{F}[f]$, hence a calculus to compute (respectively bound) Filippov
set-valued maps has been developed, see \cite{PS:87}. For example one may prove
that for a real-valued continuous function $g$ and a real-valued locally bounded
function $h$ we have $\mathcal{F}[g h](x) =
g(x)\mathcal{F}[h](x)$. In our setting this shows how to compute the Filippov
set-valued map of the right hand side $\bT$ in
\eqref{eq-de}.

\end{appendix}

\end{document}